# Adaptive AI-Driven Material Synthesis: Towards Autonomous 2D Materials Growth

*Leonardo Sabattini*[1†], *Annalisa Coriolano*[2,3†], *Corneel Casert*[4†], *Stiven Forti*[2], *Edward S. Barnard*[4], *Fabio Beltram*[1], *Massimiliano Pontil*[3], *Stephen Whitelam*[4*], *Camilla Coletti*[2], *Antonio Rossi*[2*]

[1]Scuola Normale Superiore, Laboratorio NEST, Pisa, Italy
[2]Center for Nanotechnology Innovation, Laboratorio NEST,
Istituto Italiano di Tecnologia, Pisa, Italy
[3]Computational Statistics and Machine Learning, Istituto Italiano di Tecnologia, Genova, Italy
[4]The Molecular Foundry, Lawrence Berkeley National Laboratory, Berkeley, CA, USA
† These authors contributed equally to this work: L. Sabattini, A. Coriolano, C. Casert.
* Corresponding author: antonio.rossi@iit.it, swhitelam@lbl.gov

## Abstract

Two-dimensional (2D) materials are poised to revolutionize current solid-state technology with their extraordinary properties. Yet, the primary challenge remains their scalable production. While there have been significant advancements, much of the scientific progress has depended on the exfoliation of materials, a method that poses severe challenges for large-scale applications. With the advent of artificial intelligence (AI) in materials science, innovative synthesis methodologies are now on the horizon. This study explores the forefront of autonomous materials synthesis using an artificial neural network (ANN) trained by evolutionary methods, focusing on the efficient production of graphene.

Our approach demonstrates that a neural network can iteratively and autonomously learn a time-dependent protocol for the efficient growth of graphene, without requiring pre-training on what constitutes an effective recipe. Evaluation criteria are based on the proximity of the Raman signature to that of monolayer graphene: higher scores are granted to outcomes whose spectrum more closely resembles that of an ideal continuous monolayer structure. This feedback mechanism allows for iterative refinement of the ANN's time-dependent synthesis protocols, progressively improving sample quality.

Through the advancement and application of AI methodologies, this work makes a substantial contribution to the field of materials engineering, fostering a new era of innovation and efficiency in the synthesis process.

# Introduction

The emergence of two-dimensional (2D) materials has revolutionized material science, offering promising advancements across a wide range of technological applications[1,2,3,4,5]. However, achieving scalable production of high-quality single-crystal 2D materials remains a significant challenge. Although exfoliated forms of these crystals demonstrate excellent potential for addressing numerous technological challenges, non-scalable devices made from exfoliated materials often lack reproducibility and require cumbersome fabrication methods. Achieving scalability for these materials is highly desirable, not only to facilitate practical applications but also to validate the promising scientific discoveries made to date.

Graphene with high crystallinity and charge mobility comparable to that of exfoliated graphene can be produced through chemical vapor deposition (CVD)[6,7,8]. This method was successfully demonstrated for large-scale growth, making it a promising approach for various industrial applications[9,10,11,12,13]. In contrast, the development of other 2D materials and their heterostructures still lacks a viable and scalable method. The main challenges stem from the difficulty in producing large-area crystals and, consequently, heterostructures, with controlled thickness, minimal defects, and uniformity on a large scale. As a result, the scalable synthesis of these materials has not yet matured in terms of reproducibility and defect management[14,15]. This limitation restricts the practical use and replication of promising materials across various fields such as quantum computing and sensing[16,17].

Artificial Intelligence (AI) can provide a compelling solution to this challenge. In recent years, AI has become an indispensable part of our society, proving to be extremely effective in solving complex problems across various fields, with applications ranging from precision medicine to autonomous driving[18,19,20,21,22]. In the rapidly advancing field of graphene research AI was applied to tasks such as determining the potential energy surface[23], predicting bandgaps[24], or forecasting crack evolution in graphene sheets[25]. Recent advances in AI have also been used to explore density functional theory (DFT) calculations and study diffusion mechanisms based on molecular dynamics (MD)[26,27,28,29]. One of the most striking results was shown by Merchant and coworkers who were able to predict the stability of millions of new crystal structures, many of which had never been discovered through traditional methods[30]. In this context, the integration of AI into materials science opens a new era of innovative synthesis methodologies. Recently, there has been considerable interest in establishing autonomous laboratories that combine robotics with the exploitation of ab initio databases and active learning to optimize the synthesis of novel inorganic materials[31], a process that is generally time-consuming and expensive.

In this study, we aim to address the following question: Can an Artificial Neural Network (ANN) learn a time-dependent protocol in order to grow a material with desired optical and crystalline properties, without prior knowledge of growth protocols? This is a radically different approach with respect to those mentioned so far. Here, we utilize an active learning method in which the ANN iteratively refines its synthesis protocols without relying on historical data.

Unlike methodologies used in other works[31], which rely heavily on pre-trained models and vast amounts of historical synthesis data to guide experiments, our approach distinguishes itself by requiring minimal input and by learning dynamically through direct experimental feedback. Additionally, previous methods focus on optimizing a limited state space of initial conditions and are unable to adapt to time-dependent protocols. In contrast, our ANN continuously refines its protocols over time, learning from real-time experimental outcomes and progressively improving the quality of the graphene produced.

As proof of principle, we focus on the relatively straightforward task of growing high-quality, homogeneous graphene through the thermal decomposition of silicon carbide (SiC). Here an ANN is tasked with proposing a protocol, a profile of temperature as a function of time, to achieve this goal. The growth of graphene from SiC is chosen as an ideal candidate for exploring the feasibility of applying ANN learning to crystal growth, owing to its simplicity and the manageable number of growth parameters that can be easily controlled (i.e., min temperature, max temperature and ramp). This method enables the synthesis of graphene directly from SiC[32,33], thus eliminating the need for gaseous carbon precursors.

In experimental settings, obtaining a loss function is challenging, and the gradient information required for the back-propagation algorithm is often not available. Therefore, training the neural network must rely on evolutionary methods[34,35,36]. Standard genetic algorithms are too costly for this purpose[36] since they require many runs of the experiment to train. We use instead an evolutionary algorithm that resembles the zero-temperature Metropolis Monte Carlo algorithm with an adaptive, momentum-like component, called adaptive Monte Carlo, or aMC[37]. We start with a neural network whose parameters specify a time-dependent protocol. We add random numbers from a normal distribution to all weights, evaluate the new protocol resulting from this change, and accept (or reject) the new protocol if the outcome is better (or worse) than the previous protocol, as determined by a score function. This approach proceeds iteratively, with the algorithm remembering weight changes that led to accepted protocols and making such changes with higher probability. In this way, we can achieve meaningful learning within a few tens of experiments, making the approach readily accessible to laboratory studies.

In our approach, score function evaluation relies on Raman spectroscopy measure-ments, a highly effective method extensively used for the characterization of 2D materials owing to its versatility and precision[38,39,40,41]. One of the key advantages of Raman spectroscopy is its capability to perform detailed mapping of samples, effectively distinguishing between different forms and qualities of graphene (e.g., buffer layer, monolayer graphene (MLG), and bilayer graphene (BLG)), thereby offering valuable insights into their distribution across the sample. Additionally, it is a rapid and non-invasive characterization technique that can cover an area of tens of micrometers in just a few minutes. For these reasons, it provides crucial feedback that aids in iteratively optimizing the synthesis protocols of artificial neural networks, thereby steadily improving the quality of graphene.

To further assess and validate our Raman-based scoring system, the structural, chemical, and electronic quality of graphene are characterized using supplementary techniques such as Atomic Force Microscopy (AFM), X-ray Photoelectron Spectroscopy (XPS), and Angle-Resolved Photoemission Spectroscopy (ARPES). These analyses offer comprehensive insight into the material quality, retrospectively validating the effectiveness of the AI-driven synthesis approach.

This work showcases the feasibility and significant advantages of employing artificial intelligence to tailor and optimize the growth of 2D materials. By integrating an ANN that learns and adapts, we effectively embed a "brain" into the synthesis lab, paving the way for the autonomous synthesis of desired materials. More ambitiously, this approach holds the potential to discover methods for synthesizing high-quality materials that are currently beyond our capabilities, unlocking new frontiers in material science.

# Methods

We outline here the methodology developed to train an ANN with aMC to autonomously find the most efficient growth protocol for graphene from the thermal decomposition of SiC. Minimal input is provided by setting only the furnace working temperature range and a starting temperature. The entire ANN training procedure is schematized in Figure 1 and consists of four iterative steps: (i) Protocol Generation: a temperature profile is generated by an ANN and given as an input to a cold-wall reactor; (ii) Sample growth: graphene growth on SiC is performed in the reactor adopting as input the temperature profile generated in step (i); (iii) Sample Characterization: Raman spectroscopy is performed on the synthesized sample and the spectrum obtained is benchmarked with an ideal target to generate a score (iv). At the end of this process the ANN parameters are updated following the aMC method described below, and a new protocol is generated.

Before describing the details of the aMC algorithm, we will discuss the main aspects of the growth of epitaxial graphene on SiC. It involves a high-temperature process in which the crystal is thermally decomposed[42]. The most important control parameters during growth are temperature, pressure and time. Originally this technique was implemented in ultra high vacuum (UHV) chambers[33]. It is now agreed that atmospheric-pressure growth conditions under inert gas in quartz reactors (either cold-wall or hot-wall, horizontal or vertical) provide the most favorable conditions to obtain graphene[43]. During the thermal decomposition process, silicon atoms sublimate from the surface, and leave behind a carbon-rich layer, a graphene-like honeycomb lattice with one third of the C atoms forming covalent bonds to the SiC substrate through $sp^3$-hybridized horbitals[44]. This is known as buffer or zero-layer graphene (ZLG). This layer is not yet graphene and exhibits a gap[43,45]. The growth of graphene requires reaching an optimal (higher) temperature at which an additional carbon-rich layer forms beneath the first one, effectively decoupling it from the substrate. This layer does display the typical linear dispersion of MLG[43]. If the process continues further, another buffer layer will form at the interface with SiC, turning the previous graphene and buffer layers into a BLG. Hence, determining a correct temperature profile is fundamental to obtain MLG with minimum ZLG and BLG inclusions. Given the fundamental importance of the temperature profile in achieving optimal MLG coverage, we sought to investigate a broad yet physically-meaningful temperature range. We set the temperature boundaries between $T_{min}$ = 1100°C and $T_{max}$ = 1300°C. The minimum temperature was chosen below the known temperatures for graphene synthesis to allow for a reasonably wide temperature range for the ANN to explore. The maximum temperature was set equal to the maximum operational capacity of our reactor, ensuring the respect of safety and equipment limitations. We set a starting temperature $T_{start}$ = 1200°C as midpoint within this range. This starting temperature is not too low to hinder the initiation of the growth process, yet not too close to the optimal synthesis temperature, thereby encouraging the ANN to explore a variety of temperature profiles. In subsequent runs, as discussed below, we shall remove this starting temperature constraint to allow the ANN to explore the entire temperature range more freely. Detailed experimental information about the growth process implemented in this work are reported in Supporting Information (SI).

aMC is an evolutionary algorithm designed to work with relatively few experiments: each learning step requires one experiment, and the algorithm learns from past successes, proposing similar moves with increased likelihood[37]. The algorithm proceeds as follows. Let $\boldsymbol{x}$ = $\{x_1, x_2, \ldots, x_N\}$ be the vector of neural-network parameters, which defines a time-dependent protocol. Let $U(\boldsymbol{x})$ be the loss function, which quantifies the success of the synthesis resulting from that time-dependent protocol. The algorithm proposes a change of all neural-network parameters $\mathbf{x} \to \mathbf{x}'$ by small Gaussian random numbers,

$$x_i \to x_i' = x_i + \epsilon_i, \tag{1}$$

where $\epsilon_i \sim N(\mu_i, \sigma^2)$. Here $\sigma$ sets the scale of parameter updates, and the $\mu_i$ are momentum-like parameters that we shall specify shortly. If the synthesis outcome resulting from the new protocol is better than (or as good as) the current outcome, i.e. if $U(\mathbf{x}') \leq U(\mathbf{x})$, then we accept the update (1), and the $\mathbf{x}'$ become the new parameters of the neural network. Otherwise we revert to the previous parameters. The update procedure is then repeated.

The momentum-like parameters $\mu_i$ are initially set to zero. Following each *accepted* move, they are updated as $\mu_i \to \mu_i + \eta(\epsilon_i - \mu_i)$, where $\eta$ is a hyperparameter of the method. This update ensures that subsequent parameter updates are more likely to be similar to past accepted updates. Following several consecutive *rejected* moves, momentum-like parameters are reset to zero, and the scale parameter $\sigma$ is reduced in size[37].

For the current synthesis, the loss function $U$ depends on the Raman spectrum. Each measured spectrum is assigned a score that reflects how closely it matches the desired characteristics. The loss function is inversely related to this score: a lower loss function corresponds to a higher score. The highest possible score is attributed to the ideal Raman spectrum, which serves as the benchmark for our evaluations. Consequently, the objective of the ANN training is to minimize the loss function by maximizing the similarity between the measured spectrum and the ideal Raman spectrum.

Raman spectroscopy can differentiate between ZLG, MLG, and BLG[38], offering valuable insights into their distribution across the sample. The ability to identify and map these variations in graphene layers aids in optimizing growth conditions and improving the overall quality of the 2D material. The spectrum obtained from graphene typically shows two prominent peaks: G, that is a primary in-plane vibrational mode, and 2D, the second-order peak of the in-plane vibrational mode D[46]. The intensity, position, and width of these peaks are highly sensitive to the quality and structure of graphene. High-quality graphene typically exhibits sharp and intense peaks in the Raman spectrum. For our analysis we consider the 2D peak, that can be fitted with a Lorentzian function. The score ($f$) is then calculated with the equation (2):

$$f = I + \sigma + \frac{1}{\chi}(\sigma \cdot I) \tag{2}$$

where $I$ is related to the 2D peak intensity, while $\sigma$ is related to the Lorentzian full width at half maximum (FWHM) through hyperbolic tangent, defined as follows:

$$I = Tanh\left(\frac{a \cdot I_{2D} + a}{b}\right) + 1 \tag{3}$$

and

$$\sigma = Tanh\left(\frac{a\prime \cdot FWHM + a\prime}{b\prime}\right) + 1 \tag{4}$$

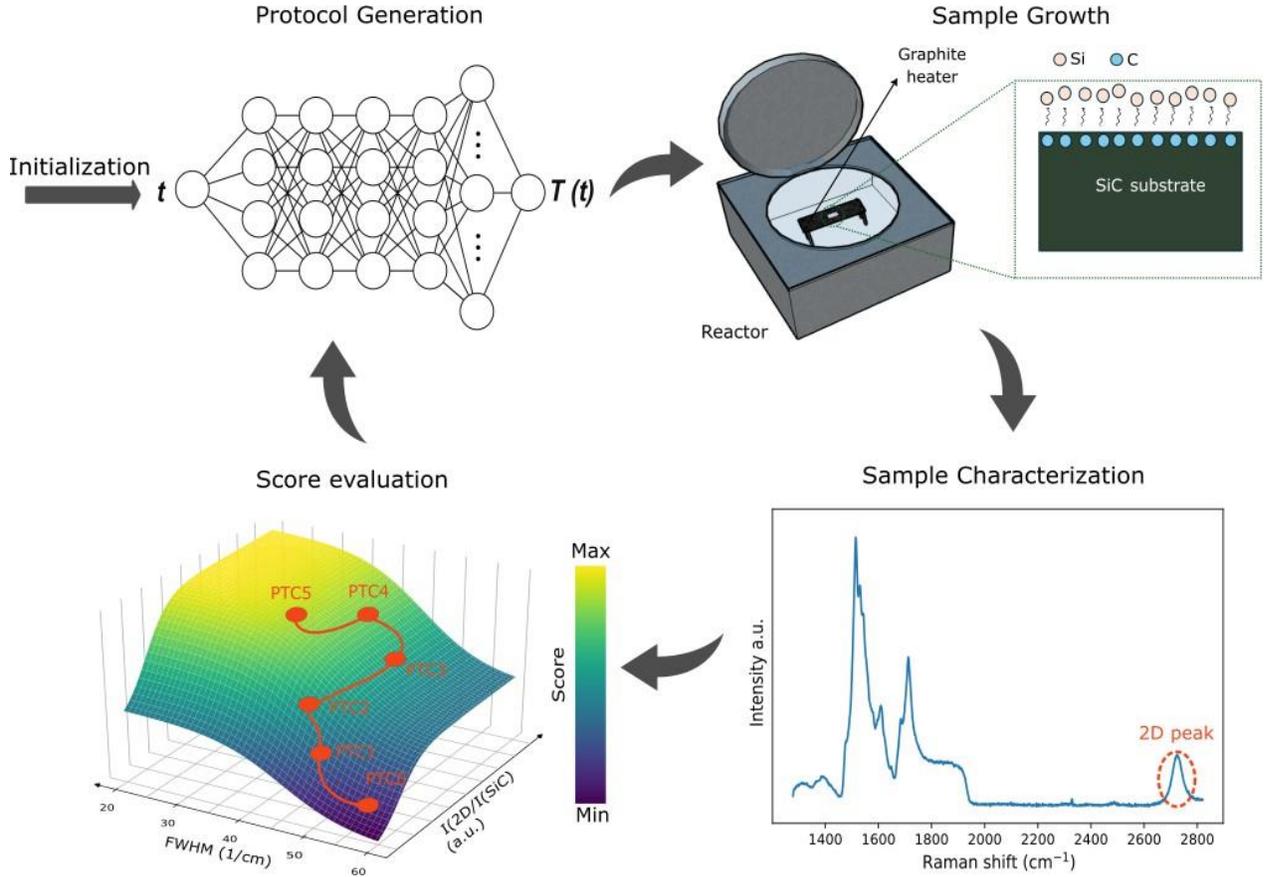

Figure 1: Schematic representation of ANN training used in this work. After initialization with the parameter guess, the protocol generation is conducted, followed by sample growth. The obtained sample is then characterized through Raman spectroscopy. The extracted data are used to evaluate the score. Finally, the protocol is updated with the new parameters.

$\chi$ is the interaction coefficient parameter, while $a$, $b$, $a'$, and $b'$ are empirical parameters.

The intensity parameter is crucial because it directly correlates with the quantity of graphene on top of the substrate, while the term correlated to the FWHM aids the protocol in distinguishing between MLG and BLG, as BLG typically exhibits a larger FWHM[47]. The choice of the hyperbolic function is dictated by the fact that beyond a threshold, the growth is complete; conversely, below a certain value, the growth is negligible. Moreover, the interference term $\frac{1}{\chi}(\sigma \cdot I)$ is necessary to prevent convergence in a region where only one of the two terms is optimized. Table S1 of SI reports the values of the empirical parameters used in (3) and (4). Figure S1 shows the 2D Raman peak (orange solid line) with the corresponding Lorentzian fit (blue solid line) for the Raman spectra corresponding to the protocols used in this work. A summary of the obtained values of $I$ and $\sigma$ are reported in Table S2. All the samples were analyzed using Raman spectroscopy under identical spectral sampling conditions. In particular, multiple samples were grown for each protocol to test the reproducibility of the process. We also checked the reproducibility of the single protocol, comparing the theoretical temperature profile with the measured one followed by the furnace (Figure S2).

# Results and discussion

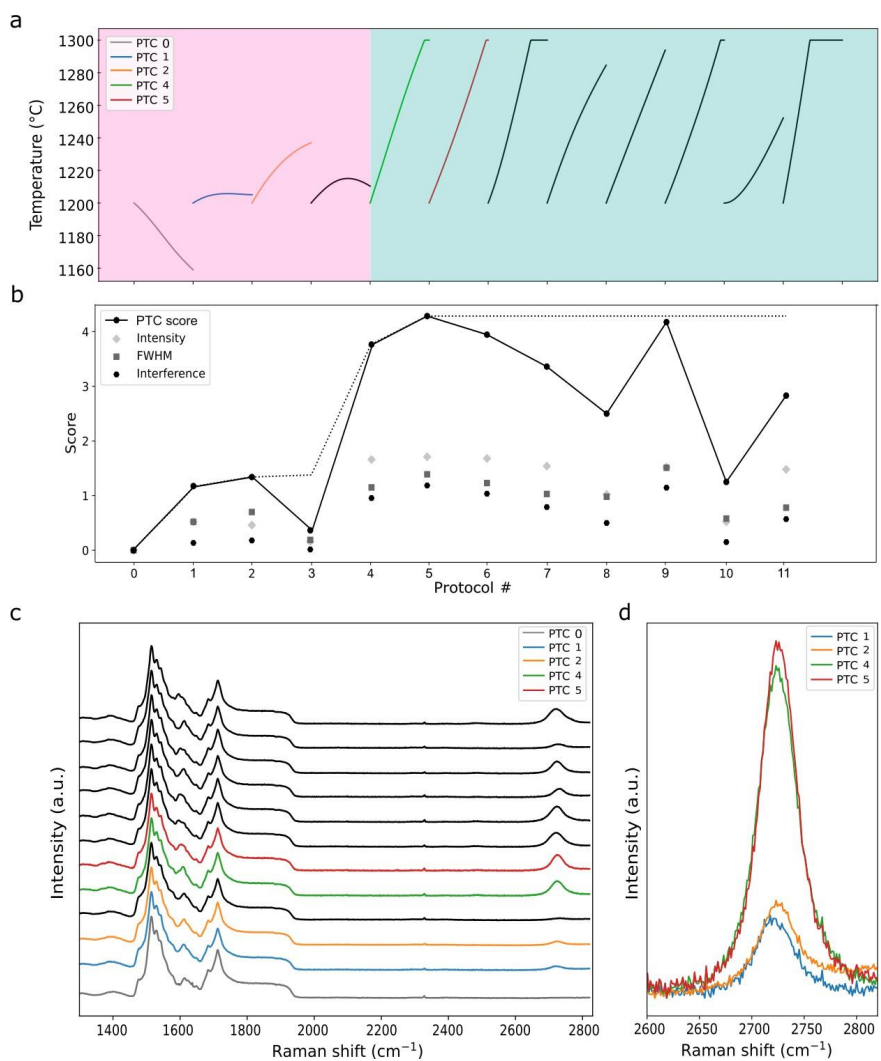

Figure 2: (a) Protocol evolution. Protocols that enhance the score are highlighted by colored lines. (b) Score evolution with the generation: The solid black line represents the score calculated using Equation 2, while the dashed black line indicates the accepted score threshold. Additionally, the intensity (diamonds), $\sigma$ (squares), and Interference parameters (circles) are depicted relative to the generation. (c) Raman spectra collected from samples used to train the ANN. Protocols that contribute to score improvement are identified by colored lines. The spectra are vertically spaced for clarity. (d) Zoomed-in Energy Range of the 2D-band Raman Peak.

The results obtained are summarized in Figure 2. The temperature profile for each generation is reported in Figure 2a. Each curve is a tentative temperature profile or protocol, proposed by the ANN and implemented into the growth set up. In Figure 2b, the solid black line represents the score calculated using Eq. 2, while the dashed black line indicates the accepted score threshold. Additionally, the intensity (diamonds), $\sigma$ (squares), and interference (circles) partial scores are shown in relation to the generation. The plots in Figure 2a-b are divided into two main regions. The first region (pale magenta) shows various attempts by the ANN to identify a protocol with a favorable score function. In contrast, the second region (pale cyan) demonstrates that the ANN has learned a successful trend, and suggests temperature profiles that are mostly monotonically increasing functions from 1200°C to the temperature upper limit (1300°C). This indicates that when the ANN encounters a protocol with a high score, it tends to keep moving in that direction. This score evolution can be visualized by analyzing the sequence of the Raman spectra reported in Figure 2c-d. At the beginning, the ANN proposes a completely random trend: for protocol (PTC) 0 (solid gray line), the temperature falls below 1200°C, which is the lower experimental limit necessary for graphene growth. This is evident from the absence of the 2D peak in the Raman spectra depicted in Figure 2c (solid gray line). In contrast, for protocols 1 and 2 (represented by the solid blue and orange lines, respectively, in Figure 2), there is an improvement: the 2D peak emerges, albeit not as intensely or sharply as in high-quality graphene growth. Subsequently, through additional generations, the ANN further enhances the graphene quality, as evidenced by protocols 4 and 5 (indicated by the solid green and red lines, respectively, in Figure 2). In these instances, a distinct and sharp 2D peak is observed. Figure 2d displays a zoomed-in energy range of the 2D Raman Peak for the protocols that yield improved scores, normalized with respect to the SiC peak. It's noteworthy that while the intensity value of the 2D peak for protocol 4 closely resembles that of protocol 5, the score for protocol 4 consistently falls below that of protocol 5 (Solid black line in Figure 2b). This discrepancy can likely be attributed to the broader FWHM of the 2D peak in protocol 4, indicating the presence of a BLG. Thus, this reaffirms the significance of the interference term within the score evaluation formula. ANN training is also conducted by removing the lower limit temperature constraint (see Figure S3). It is evident that, even lifting this constrain, the ANN successfully learns the desired patterns in a similar manner with a slightly higher number of protocols. Initially, the temperature profiles generated by the ANN start at T lower than 1200°C. However, over time, the ANN adapts and begins to suggest temperature profiles that correspond to effective graphene synthesis. This adaptation is noticeable, as in the previous run, in the pale red region of the plot, where the suggested temperature profiles align with known effective growth conditions.

To visualize the learning process and validate our scoring mechanism, we conducted a series of cross-checked experiments using AFM, XPS and ARPES, providing a comprehensive set of surface characterization tools. Starting with AFM, adhesion force maps are shown in Figure 3a-d. Adhesion AFM depends on the interactions between the probe tip and the sample surface, which can be influenced by factors such as surface roughness, chemical composition, and the presence of contaminants or adsorbates[48]. Hence, it can serve as a valuable method to quantify the amount and quality of graphene on the surface.

The adhesion force is strictly correlated to the material Young's modulus[49]. The darkest areas (i.e., low adhesion force) represent the SiC surface, while the lightest area (i.e., high adhesion force) represents the graphene surface. It is possible to note that with the score improvement there is an increase of the graphene area from 22.4% (protocol 1, Figure 3a) to 88.2% (protocol 5, Figure 3d). For protocol 4, the adhesion map shows a high percentage coverage of graphene. However, the relative adhesion force values are less distinct. Generally, the Young's modulus of MLG is higher than that of BLG[50], thus providing an explanation for the different adhesion contrast and confirming the presence of BLG on the sample obtained from protocol 4. The percentage of the graphene surface as a function of the protocol number is reported in Figure S4.

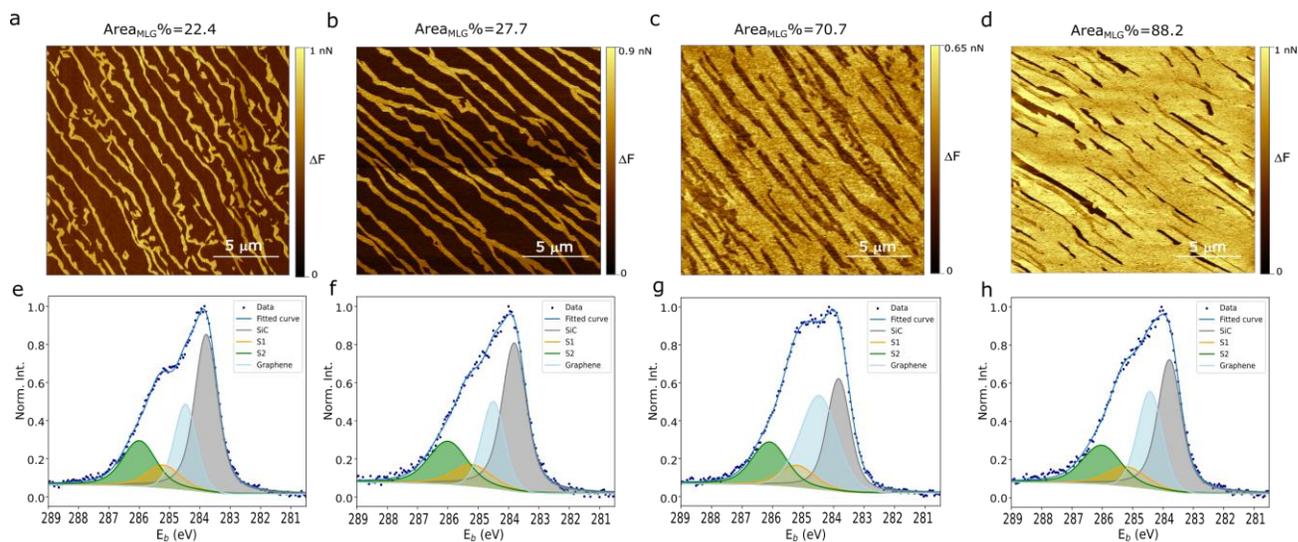

Figure 3: (a-d) Adhesion force map on samples obtained from PTC1 (a), PTC2 (b), PTC4 (c) and PTC5 (d). (e-h) C1s XPS spectra recorded on samples obtained from PTC1 (e), PTC2 (f), PTC4 (g) and PTC5 (h).

The AFM analysis is cross-checked looking at the chemical properties of the samples measuring the core-level spectrum of each sample via XPS. The results are presented in Figure 3e-h. The spectral intensities are normalized to facilitate comparison among the different samples. We found that the $sp^2$ position peak occurs at around 284.4 ± 0.1 eV. The SiC component is identified at 283.8 ± 0.1 eV, while for the components associated with the buffer layer, S1 and S2, the peaks are centered at 285.2 ± 0.1 eV and 286.0 ± 0.1 eV, respectively, in good agreement with the literature[32,51]. The core-level fitting procedure used in this work is detailed in the SI, "XPS fitting procedure" section (see Figure

S5 and Table S3). The area subtended by the graphene curve is 18.94% for protocol 1 (Figure 3e), 19.60% for protocol 2 (Figure 3f), 31.30% for protocol 4 (Figure 3g), and 23.10% for protocol 5 (Figure 3h), also observing in this case an improvement in the quality of the graphene with the increase in the score. It is important to note that the FWHM of the graphene curve for protocol 4 is slightly higher compared to the other protocols, due to the Van der Waals interactions between two graphene layers, further confirming the presence of BLG in the sample obtained through protocol 4 in great agreement with the AFM analysis. In Figure S6 we report the trend of the percentage of the graphene area obtained from the XPS fit as a function of the protocol number. The dashed black line indicates the threshold of the area for a fully covered sample. This means that for a percentage of area higher than approximately 26%, there is the presence of BLG on the substrate, as shown in the case of the sample obtained with PTC4.

Finally, the band structure of the as-grown samples is investigated using ARPES (see Figure 4). This technique is ideally suited to probe the electronic band structure of a material and can provide a definitive assessment of graphene quality. The spectra are collected at the $\overline{K}$-point of the graphene Brillouin zone, along the $\overline{\Gamma K}$ direction in reciprocal space. We use the sharpness of the graphene bands as a metric to quantify the quality of the relative protocol. We fit the momentum distribution curves (MDC), extracted 50 meV below the Fermi level (colored dashed lines in Figure 4a-d), reported in the bottom panels of Figure 4: the FWHM ($2\gamma$) of the Voigt function used to fit the peaks gradually decreases from 0.044 Å$^{-1}$ (protocol 1, Figure 4a) to 0.020 Å$^{-1}$ (protocol 5, Figure 4d). As expected, the ARPES spectrum collected on the sample obtained from protocol 4, displayed in Figure 4c, shows the classic band dispersion of a BLG on SiC[52], overlapping to the MLG bands, with a splitting of the $\pi$ band originated from the interlayer interaction between the two graphene layers.

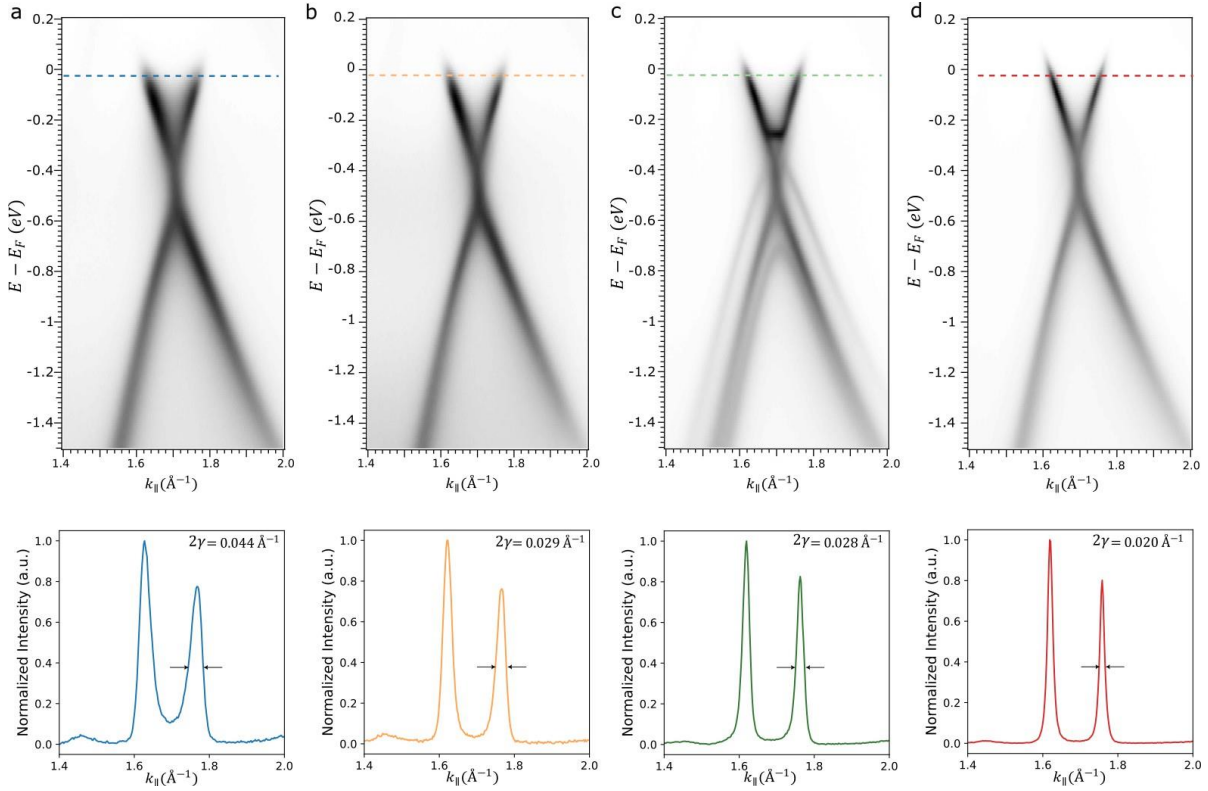

Figure 4: (Top Panels) ARPES intensity maps collected on samples obtained from PTC1 (a), PTC2 (b), PTC4 (c) and PTC5 (d). (Bottom Panels) The corresponding normalized MDC spectra, obtained by integrating the signal at the Fermi level energy (indicated with colored dashed lines in the ARPES maps).

These supplementary chemical and structural analyses show a trend that align closely with the calculated score function, validating the use of Raman characterization for the score function in ANN training.

## Conclusions

We have demonstrated the potential for autonomous synthesis by using an adaptive learning algorithm to train an artificial neural network that encodes a time-dependent synthesis protocol. The neural network has iteratively and autonomously learned to synthesize high-quality graphene, with minimal initial input, thus demonstrating its remarkable ability to learn and adapt. This capability resulted in progressively improved graphene quality, verified through comprehensive surface characterization techniques. The success of this approach highlights the ANN's robust learning mechanisms and adaptability, underscoring its capacity to handle complex material synthesis tasks. By embedding intelligent decision-making into the synthesis process, our work paves the way for future applications of AI in the growth of various 2D materials. This advancement not only promises significant efficiency and quality improvements in material production but also opens new frontiers in material science, where AI-driven methods can explore and optimize previously unattainable synthesis pathways.

## Conflicts of interest

There are no conflicts to declare.

## Acknowledgements

We acknowledge the project PNRR MUR Project PE000013 CUP J53C22003010006 "Future Artificial Intelligence Research (FAIR) and PNRR MUR Project PE0000023 - National Institute of Quantum Science and Technology (NQSTI) funded by the European Union - NextGenerationEU.

*Physics*, vol. 47, no. 9, p. 094013, 2014.

# Supporting Information: Adaptive AI-Driven Material Synthesis: Towards Autonomous 2D Materials Growth

*Leonardo Sabattini*[1†], *Annalisa Coriolano*[2,3†], *Corneel Casert*[4†], *Stiven Forti*[2], *Edward S. Barnard*[4], *Fabio Beltram*[1], *Massimiliano Pontil*[3], *Stephen Whitelam*[4], *Camilla Coletti*[2], *Antonio Rossi*[2]

[1]Scuola Normale Superiore, Laboratorio NEST, Pisa, Italy
[2]Center for Nanotechnology Innovation, Laboratorio NEST,
Istituto Italiano di Tecnologia, Pisa, Italy
[3]Computational Statistics and Machine Learning, Istituto Italiano di Tecnologia, Genova, Italy
[4]The Molecular Foundry, Lawrence Berkeley National Laboratory
† These authors contributed equally to this work: L. Sabattini, A. Coriolano, C. Casert. * Corresponding author: antonio.rossi@iit.it, swhitelam@lbl.gov

## Experimental details

Epitaxial graphene is grown by thermal decomposition starting from SiC substrate. Commercial SiC wafers, thoroughly cleaned with acetone and IPA in an ultrasonic bath, are subsequently treated with oxygen plasma, followed by Piranha solution and HF baths, to remove all organic contaminants from the surface. Subsequently, the SiC (0001) wafers are exposed to high-temperature hydrogen gas within a furnace to remove polishing scratches, a process known as hydrogen etching[1]. Finally, the graphene growth process starts. All growth procedures are conducted using an AIXTRON Black Magic cold-wall reactor in an Ar/H$_2$-filled chamber

Raman spectra are acquired using a Renishaw InVia Raman microscope with a con-tinuous wave 532 nm laser using a 100x objective and 3.25 mW incidence power. For each sample, three 10x10 $\mu m^2$ maps are collected with a step size of 1 $\mu$m and an exposure time of 5 seconds each. The AFM images are acquired with a Bruker Dimension Icon microscope operated in quantitative nanomechanical mapping (QNM) mode in ambient conditions. Photoemission data are aquired with a hemispherical analyzer (SPECS PHOIBOS 150). X-ray photoelectron spectroscopy (XPS) spectra are obtained with a SPECS XR-50 Al K$\alpha$ X-ray source. Angle resolved photoemission spectroscopy (ARPES) are collected using He I$\alpha$ radiation (21.2 eV) excitation source (SPECS-$\mu$Sirius) with a nominal spot size of 100 $\mu m$.

## Score Function Evaluation

The empirical parameters *a, b, a′* and *b′* used in the equations (3) and (4) of the main text to calculate the score function, are obtained based on a previous work[2] and our previous tests. The values of the parameters are reported in Table S1.

| a | b | a′ | b′ |
|---|---|---|---|
| -3 | 3 | -45 | 10 |

Table S1: Score parameters.

The fitting process is Figure S1 shows the 2D Raman peak (orange solid line) with the corresponding Lorentzian fit (blue solid line) for the Raman spectra corresponding to the protocols used in this work.

Table S2 summarizes the values used to calculate the score evolution reported in Figure 2b of the main text.

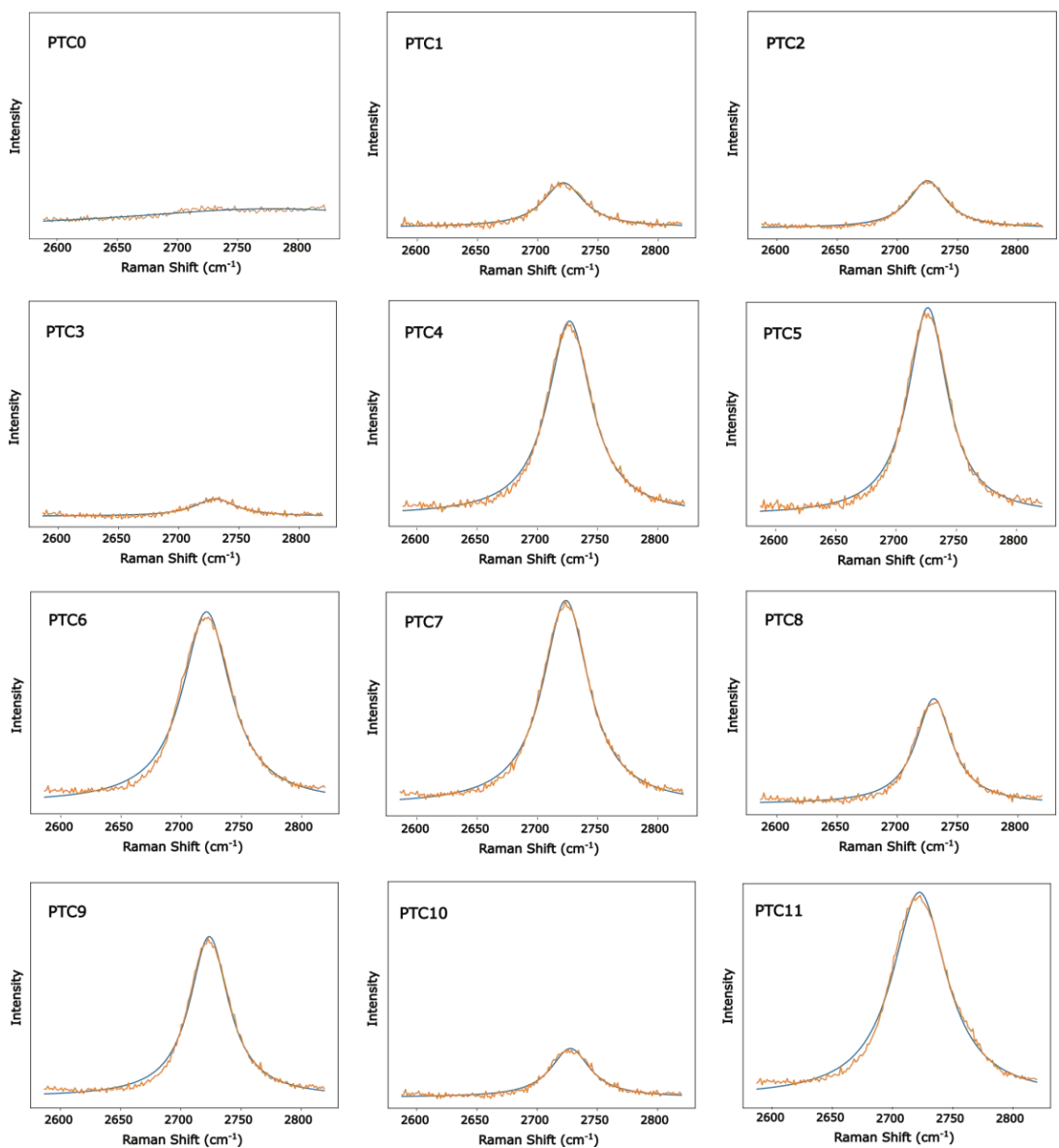

Figure S1: 2D Raman peak with the corresponding Lorentzian fit for the Raman spectra of the sample used in this work.

| PTC # | I | $\sigma$ | $\frac{1}{x}(\sigma \cdot I)$ |
|---|---|---|---|
| 0 | 0.00 | 0.00 | 0.00 |
| 1 | 0.52 | 0.52 | 0.14 |
| 2 | 0.46 | 0.70 | 0.18 |
| 3 | 0.16 | 0.19 | 0.02 |
| 4 | 1.66 | 1.15 | 0.96 |
| 5 | 1.71 | 1.39 | 1.19 |
| 6 | 1.68 | 1.23 | 1.04 |
| 7 | 1.54 | 1.03 | 0.79 |
| 8 | 1.02 | 0.98 | 0.5 |
| 9 | 1.52 | 1.51 | 1.15 |
| 10 | 0.52 | 0.58 | 0.15 |
| 11 | 1.48 | 0.78 | 0.57 |

Table S2: Summary of the values used to calculate the score function.

## Graphene Growth

Figure S2 shows the theoretical temperature profile as a function of time (red line) and the experimental temperature profiles followed by the furnace for three different samples obtained through the same protocol.

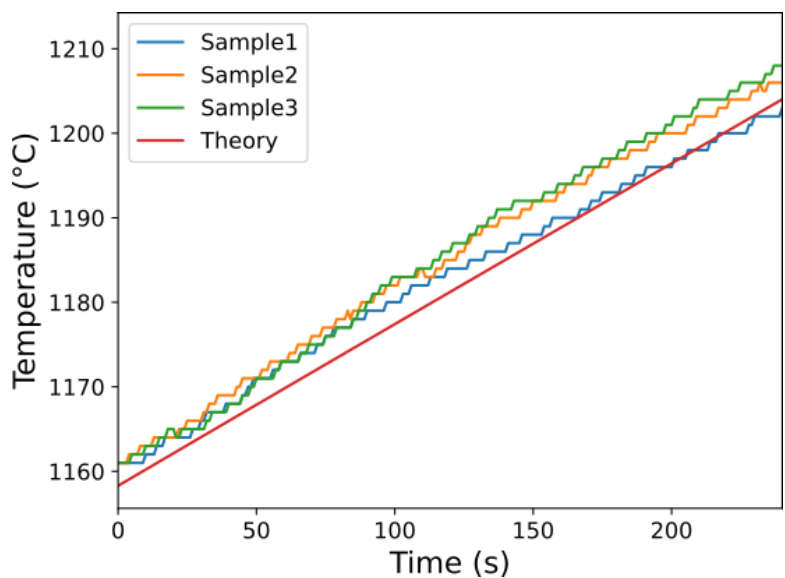

Figure S2: Theoretical (red line) and experimental profiles of temperature in function of time for different samples obtained through the same protocol.

## Temperature constrain

In Figure S3 is shown the ANN training process obtained by removing the lower temperature constrain. It is evident that, even in this scenario, the ANN has successfully learned the optimal growth protocol trend. Initially, the protocols generated start at temperatures lower than 1200°C. Over time, the ANN begins to suggest temperature profiles that align with effective graphene production, as seen in the pale red region of the plot.

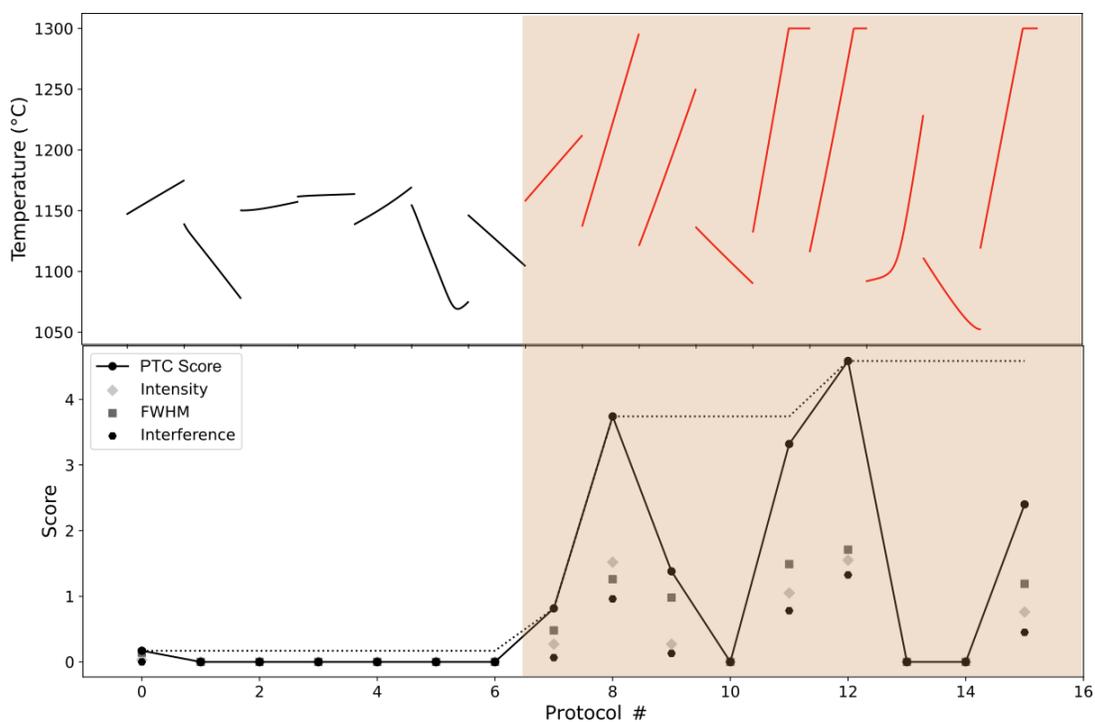

Figure S3: Evolution of the temperature profiles (upper panel) and of the score (lower panel) obtained by removing the lower temperature limit during ANN training.

# AFM measurements

Figure S4 shows the percentage of the graphene surface as a function of the protocol number, extracted from the relative adhesion maps reported in Figure 3a-d of the main text.

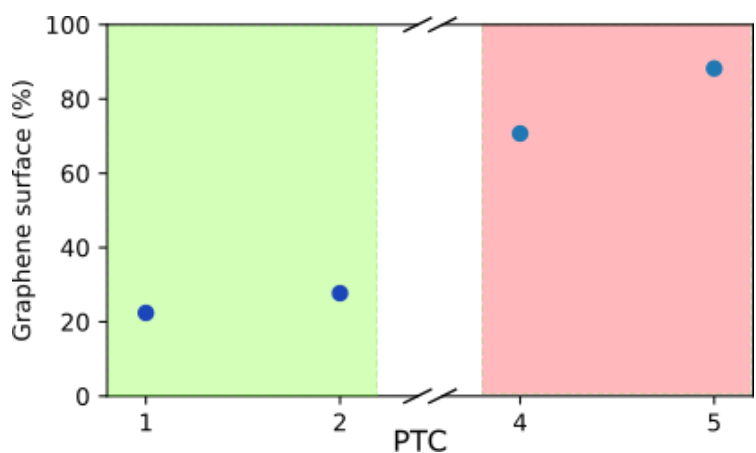

Figure S4: Percentage of the graphene surface as a function of the protocol number.

# XPS fitting procedure

For each XPS spectrum presented in Figure 3e-h, a Shirley-type background is considered. Symmetric peaks are modeled using Voigt functions (SiC, S1 and S2, where S1 and S2 represent the components associated with the buffer layer), while for the graphene peak, to address the asymmetry arising from conductive layers, a Gaussian-Doniach-Šunjić (GDS) line shape is employed[3,4]. In particular, two additional components (denoted S1$^I$ and S2$^I$) are considered for BL, to also account for the parts of BL that interact with graphene. Figure S5 shows the core-level fitting with the additional BL components, which have been removed from Figure 3 in the main text for clarity.

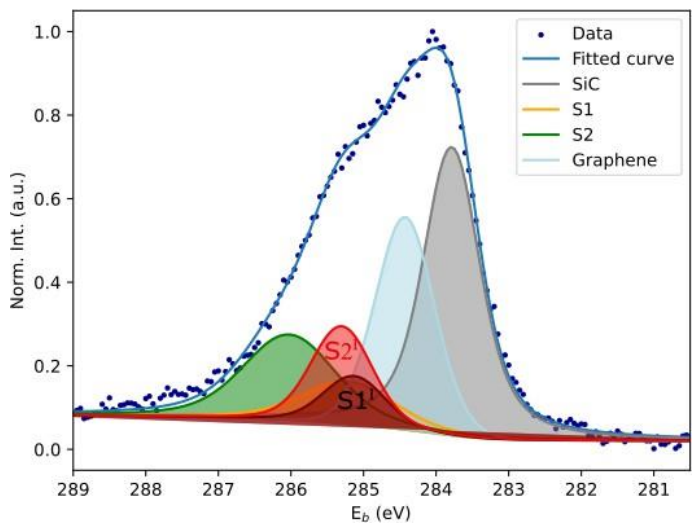

Figure S5: Core-level fitting of XPS spectrum collected on sample obtained from PTC5 showing also the additional BL components.

Table S3 summarizes the values for the graphene curve extracted from the fitting procedure, where $\omega_G$ denotes the FWHM for the Gaussian component, while $\omega_L$ denotes the FWHM for the Lorentzian component.

| PTC # | $A_{graphene}\%$ | Peak position (eV) | $\omega_G$ (eV) | $\omega_L$ (eV) |
|---|---|---|---|---|
| 1 | 18.94 | 284.45 | 1.15 | 1.38 |
| 2 | 19.60 | 284.45 | 1.13 | 1.36 |
| 4 | 31.30 | 284.40 | 1.34 | 1.38 |
| 5 | 23.10 | 284.40 | 1.13 | 1.36 |

Table S3: Summary of the values for the graphene curve extracted from the fitting procedure.

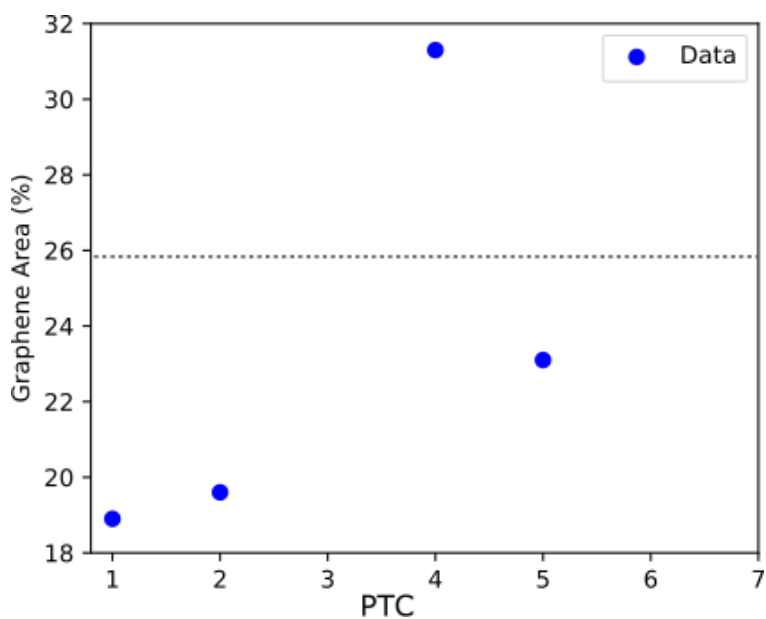

Figure S6: Percentage of the graphene area obtained from the XPS fit as a function of the protocol number. The dashed black line indicates the threshold of the area for a fully covered sample.

In Figure S6 we report the trend of the percentage of the graphene area obtained from the XPS fit as a function of the protocol number. The dashed black line indicates the threshold of the area for a fully covered sample.